\documentclass[12pt]{article}
\usepackage{ctex}
\usepackage{amssymb,amsmath,color,mathdots}
\usepackage[colorlinks,linkcolor=blue]{hyperref}
\usepackage{geometry}
\usepackage{tabularx}
\usepackage{longtable}
\usepackage{caption}
\usepackage{threeparttable}

\usepackage{bm}
\usepackage{authblk}
\allowdisplaybreaks[4]
\usepackage[square, comma, sort&compress, numbers]{natbib}
\usepackage{enumerate}

\begin{document}
	\title{\bf On the Ding and Helleseth's 8th open problem about optimal ternary cyclic codes}
	\author{\small Dong He, Peipei Zheng}	
	\author{\small Qunying Liao
		\thanks{Corresponding author.
			
			{~~E-mail. qunyingliao@sicnu.edu.cn (Q. Liao), 759803613@qq.com(D. He).}
			
			{~~Supported by National Natural Science Foundation of China (12471494) and Natural Science Foundation of Sichuan Province (2024NSFSC2051).} }
	}
	\affil[] {\small(College of Mathematical Science, Sichuan Normal University, Chengdu, Sichuan, 610066, China)}
	\date{}
	\maketitle
	\newtheorem{definition}{Definition}[section]
	\newtheorem{lemma}{Lemma}[section]
	\newtheorem{theorem}{Theorem}[section]
	\newtheorem{proposition}{Proposition}[section]
	\newtheorem{corollary}{Corollary}[section]
	\newtheorem{remark}{Remark}[section]
	\renewcommand\refname{References}	
	\renewcommand{\theequation}{\thesection.\arabic{equation}}
	\newcommand{\upcite}[1]{\textsuperscript{\textsuperscript{\cite{#1}}}}
	\catcode`@=11 \@addtoreset{equation}{section} \catcode`@=12
	{\bf Abstract.}  {\small The cyclic code is a subclass of linear codes and has applications in consumer electronics, data storage systems and communication systems due to the efficient encoding and decoding algorithms. In 2013, Ding, et al. presented nine open problems about optimal ternary cyclic codes. Till now, the 1st, 2nd, 6th and 7th problems were completely solved, the 3rd, 8th and 9th problems were incompletely solved. In this manuscript, we focus on the 8th problem. By determining the root set of some special polynomials over finite fields, we present a counterexample and a sufficient condition for the ternary cyclic code $\mathcal{C}_{(1, e)}$ optimal. Furthermore, basing on the properties of finite fields, we construct a class of optimal ternary cyclic codes with respect to the Sphere Packing Bound, and show that these codes are not equivalent to any known codes.} \\

	\section{Introduction}
	Let $p$ be a prime and $m$ be a positive integer. Let $\mathbb{F}_p$ and $\mathbb{F}_{p^m}$ denote the finite fields with $p$ and $p^m$ elements, respectively. A linear code $\mathcal{C}$ with parameters $[n, k, d]$ over the finite field $\mathbb{F}_p$ is a $k$-dimensional subspace of $\mathbb{F}_p^n$ with minimum Hamming distance $d$. $\mathcal{C}$ is a cyclic code if any cyclic shift of a codeword is also a codeword in $\mathcal{C}$. For the case $\operatorname{gcd}(n, p)=1$, a cyclic code $\mathcal{C}$ can be expressed as $\mathcal{C}=\langle g(x)\rangle$, where $g(x)$ is monic. $g(x)$ is called the generator polynomial of $\mathcal{C}$ and $h(x)=(x^n-1) / g(x)$ is called the parity-check polynomial of $\mathcal{C}$. The cyclic code is a  class of linear codes with applications in both communication systems and consumer electronics. The recent advances and contributions can be found in a series of publications including  \cite{A1,A3,A22,A25, A4,A5,A6,A8,A7,A18,A24}, as well as the references therein.
	
	Let $\alpha$ be a generator of $\mathbb{F}_{p^m}^*=\mathbb{F}_{p^m} \backslash\{0\}$ and $m_i(x)$ be the minimal polynomial of $\alpha^i$ over $\mathbb{F}_p$, where $1 \leq i \leq p^m-1$.  The cyclic code over $\mathbb{F}_p$ with generator polynomial $m_u(x) m_v(x)$ is denoted by $\mathcal{C}_{(u, v)}$, where $u$ and $v$ are from the different $p$-cyclotomic cosets. When $p=3$, the ternary cyclic code with parameters $[3^m-1,3^m-1-2 m, 4]$ is distance-optimal with respect to the Sphere Packing Bound \cite{A9}. For the case $u \neq 1$, several classes of optimal ternary cyclic codes $\mathcal{C}_{(u, v)}$ have been proposed \cite{A18,A19,A5,A23}. For the case $u = 1$, Carlet et al. \cite{A10} proposed several optimal ternary cyclic codes basing on perfect nonlinear monomials over $\mathbb{F}_{3^m}$. In 2013, Ding et al. \cite{A11} constructed some new classes of optimal ternary cyclic codes by using almost perfect nonlinear monomials（APN） over $\mathbb{F}_{3^m}$ and  some other monomials. Moreover, Ding et al. presented nine open problems by using the monomial $x^v$ over $\mathbb{F}_{3^m}$. Till now, the 1st, 2nd and 6th problems were completely solved \cite{A12,A13,A14}. Zha et al. \cite{A17} considered a special case of the 3rd problem and obtained some new optimal ternary cyclic codes. Recently, Ye et al. \cite{A21}  presented a counterexample and gave an incomplete answer for the 7th problem.  And the 8th and 9th problems for some special $h$ were studied \cite{A15,A16}.
	
	In this manuscript, we present a counterexample for the 8th problem and give two classes of optimal ternary cyclic codes by checking the conditions $Q_1$, $Q_2$ and $Q_3$ in Lemma \ref{l14}. The manuscript is organized as follows. In Section \ref{h1}, we introduce some preliminaries needed. In Section \ref{h2}, we present a counterexample and give a class of optimal ternary cyclic codes with parameters $[3^m-1,3^m-1-2 m, 4]$ by determining the root set of some special polynomials over finite fields. In Section 4.1, we give a class of optimal ternary cyclic codes $\mathcal{C}_{(1, e)}$ with parameters $[3^m-1,3^m-1-2 m, 4]$ basing on the properties of finite fields. In Section 4.2, we show that our new optimal cyclic codes are not equivalent to any known codes. In Section \ref{h4}, the conclusion of the whole manuscript is given.

	\section{Preliminaries}\label{h1}
	In this section, we first introduce the $p$-cyclotomic coset.
	Let $n=p^m-1$, for any integer $i$ with $0 \leq i \leq n-1$, the $p$-cyclotomic coset modulo $n$ containing $i$ is defined by
	$$
	C_i=\{i p^s ~(\bmod~n) \mid s=0,1, \ldots, \ell_i-1\}
	,$$
	where $\ell_i$ is the minimal positive integer such that $p^{\ell_i} i \equiv i~(\bmod~n)$, and $\ell_i$ is the size of $C_i$, denoted by $|C_i|$.
	
	The following lemmas are necessary.
	
		\begin{lemma}\cite[Lemma 2.1]{A11}\label{l11}
		For any integer $e$ with $1 \leq e \leq 3^m-2$ and $\operatorname{gcd}\left(e, 3^m-1\right)=2$, we have $\left|C_e\right|=m$.
	\end{lemma}
	
	\begin{lemma}\cite[Theorem 3.46]{A20}\label{l12}
		Let $k$ be a positive integer and $f$ be an irreducible polynomial  of degree $l$ over $\mathbb{F}_p$. Then $f$ can be factorized into $d$ irreducible polynomials in $\mathbb{F}_{p^{k}}[x]$ of the same degree $\frac{l}{d}$, where $d=\operatorname{gcd}(k, l)$.
	\end{lemma}
	
		\begin{lemma}\cite[Lemma 2.1]{A15}\label{l13}
		For any integer $i$ with $0 \leq i \leq n-1$, we have $\ell_i \mid m$, where $\ell_i$ is the size of $C_i$.
	\end{lemma}
	
	The ternary cyclic code $\mathcal{C}_{(1, e)}$ with parameters $[3^m-1,3^m-1-2 m, 4]$ is optimal with respect to the Sphere Packing Bound, and the following sufficient and necessary conditions were given by Ding and Helleseth \cite{A11}. 
	\vspace{3mm}
	\begin{lemma}\cite[Theorem 4.1]{A11}\label{l14}
		Let $e \notin C_1$ and $|C_e|=m$. Then the ternary cyclic code $\mathcal{C}_{(1, e)}$ has parameters $[3^m-1,3^m-1-2 m, 4]$ if and only if the following conditions are satisfied simultaneously:
		
		$Q_1$. $e$ is even;
		
		$Q_2$. the equation $(x+1)^e+x^e+1=0$ has the unique solution $x=1$ in $\mathbb{F}_{3^m}$;
		
		$Q_3$. the equation $(x+1)^e-x^e-1=0$ has the unique solution $x=0$ in $\mathbb{F}_{3^m}$.
	\end{lemma}

	\section{The first class of optimal ternary cyclic codes with minimum distance four}\label{h2}
	In this section, we give a class of optimal ternary cyclic codes $\mathcal{C}_{(1, e)}$ with respect to the Sphere Packing Bound, which is an incomplete answer for the 8th problem in \cite{A11}.
	
	\vspace{0.8em}
	\noindent {\bf The 8th Open Problem} \cite{A11}\quad
	Let $m $ be an odd prime. Let $e=3^h+13$, where $3 \leq h \leq m-1$. What are the conditions on $h$ under which the ternary cyclic code $\mathcal{C}_{(1, e)}$ has parameters $[3^m-1,3^m-1-2 m, 4]$ ?
	
	\vspace{0.8em} Before give our main results, we first present a counterexample for the above problem as follows. 
	$\\[8pt]$
	{\bf Example 3.1}\quad Let $m=773$ be an odd prime, $h=7$ and $e=3^h+13=2200$. Basing on Magma program, we can factorize $(x+1)^e-x^e-1$ into the product of the irreducible polynomials over $\mathbb{F}_3$ as follows,\\
	$	(x+1)^{2200}-x^{2200}-1=
	(x^{773} + x^{771} + x^{769} -x^{768} + x^{767} + x^{765} + x^{763} + x^{760} + x^{759} + x^{757} + x^{756} + x^{754} + x^{753} + x^{752} -x^{751} + x^{750} + x^{749} + x^{748} + x^{747} -x^{746} + x^{744} -x^{743} + x^{742} + x^{741} + x^{740} + x^{739} -x^{738} +
	x^{736} + x^{734} -x^{732} + x^{731} + x^{730} + x^{729} -x^{728} + x^{727} + x^{725} -x^{723} + x^{721} + x^{720} + x^{718} -x^{717} + x^{716} -x^{715} + x^{714} -x^{713} + x^{712} + x^{711} -x^{710} + x^{709} -x^{708} -x^{707} -x^{706} +
	x^{705} + x^{703} -x^{702} -x^{701} -x^{699} + x^{698} -x^{697} + x^{695} -x^{692} -x^{691} -x^{689} -x^{686} +
	x^{685} -x^{684} + x^{683} -x^{681} -x^{680} + x^{679} + x^{677} -x^{675} + x^{674} + x^{673} -x^{671} -x^{670} +
	x^{669} + x^{667} + x^{666} + x^{665} -x^{663} -x^{662} -x^{661} + x^{660} -x^{659} + x^{658} + x^{657} -x^{656} -x^{655} -x^{654} -x^{653} + x^{650} -x^{649} -x^{646} + x^{645} + x^{640} + x^{639} -x^{636} -x^{635} -x^{634} -x^{633}+ x^{632} + x^{631} -x^{630} + x^{629} -x^{628} + x^{626} + x^{625} -x^{623} -x^{622} + x^{621} + x^{620} -x^{619} + x^{614} + x^{613} + x^{612} -x^{611} -x^{610} -x^{609} -x^{608} -x^{607} + x^{606} -x^{603} + x^{600} -x^{597} + x^{596} + x^{595} -x^{593} + x^{592} + x^{590} -x^{589} + x^{588} -x^{587} -x^{586} + x^{584} -x^{582} +
	x^{581} + x^{580} -x^{578} -x^{574} + x^{572} -x^{569} + x^{568} -x^{565} + x^{564} + x^{562} + x^{561} -x^{560} + x^{558}
	-x^{557} + x^{556} + x^{555} + x^{554} -x^{553} + x^{548} -x^{547} + x^{546} -x^{545} -x^{544} -x^{543} + x^{542} -x^{541} + x^{540} -x^{538} -x^{537} -x^{536} -x^{535} + x^{534} + x^{533} -x^{530} -x^{529} + x^{526} + x^{525} -x^{524} + x^{522} -x^{520} + x^{519}+ x^{518} -x^{517} -x^{516} + x^{513} -x^{512} -x^{511} + x^{510} + x^{507} -x^{506} + x^{505} -x^{502}+ x^{499} -x^{496} + x^{494} + x^{492} + x^{489} + x^{487} -x^{486} -x^{484} -x^{483} -x^{482} + x^{481} -x^{480} -x^{479} -x^{478} -x^{477 }+ x^{474} + x^{473} + x^{472} + x^{468} -x^{467} -x^{466} +
	x^{465} + x^{464} + x^{463} + x^{462} -x^{461} + x^{460} + x^{458} -x^{457} + x^{456} -x^{455} + x^{454} + x^{452} -x^{450} +
	x^{449} + x^{448} -x^{446} -x^{445} + x^{444} + x^{443} -x^{442} -x^{441} -x^{439} -x^{437} + x^{435} + x^{434} +
	x^{433} + x^{432} + x^{431} + x^{429} + x^{426} + x^{425} -x^{424} + x^{422} -x^{421} -x^{420} -x^{419} -x^{418} -x^{415} -x^{414} + x^{412} -x^{410} + x^{407} -x^{405} + x^{404} + x^{403} + x^{400} + x^{399} -x^{398} + x^{397} -x^{395} -x^{394} + x^{392} + x^{391} -x^{390} + x^{389} + x^{388} -x^{387} + x^{386} + x^{384} -x^{382} -x^{381} -x^{380} -x^{379} + x^{376} -x^{375} -x^{373} -x^{372} -x^{371} + x^{369} + x^{368} + x^{365} + x^{364} + x^{363} +
	x^{361} + x^{360} -x^{358} -x^{357} -x^{355} + x^{354} -x^{353} -x^{352} -x^{350} + x^{349} + x^{348} + x^{347} +
	x^{346} + x^{343} -x^{342} + x^{341} + x^{339} + x^{337} -x^{334} -x^{333} -x^{332} -x^{331} -x^{330} + x^{329} +
	x^{328} + x^{326} + x^{325} -x^{323} -x^{322} -x^{320} + x^{319} -x^{318} -x^{317} + x^{315} + x^{313} -x^{312} -x^{310} + x^{309} -x^{308} -x^{306} -x^{305} + x^{304} -x^{302} + x^{300} + x^{296} -x^{295} + x^{293} -x^{291} -x^{288} + x^{287} -x^{286} + x^{285} + x^{284} + x^{283} -x^{281} -x^{280} -x^{278} -x^{275} + x^{274} + x^{270} -x^{267} -x^{265} -x^{264} -x^{259} + x^{258} + x^{257} -x^{255} + x^{254} + x^{250} + x^{249} + x^{244} -x^{243} +
	x^{242} + x^{241} + x^{240} -x^{239} + x^{238} -x^{235} -x^{234} + x^{232} -x^{229} -x^{228} -x^{227} + x^{226} -x^{225} + x^{224} + x^{223} + x^{222} + x^{221} -x^{220} -x^{219} -x^{217} -x^{216} -x^{215} + x^{214} + x^{213} -x^{212} + x^{211} + x^{210} -x^{209} -x^{208} + x^{207} + x^{205} -x^{204} + x^{202} -x^{200} + x^{198} + x^{197} + x^{195}
	-x^{194} + x^{193} -x^{192} + x^{190} -x^{188} + x^{186} -x^{183} -x^{182} + x^{181} + x^{179} -x^{178} + x^{177} -x^{175} + x^{174} + x^{171} -x^{169} -x^{165} + x^{164} + x^{163} -x^{162} + x^{161} + x^{160} + x^{159} -x^{158} -x^{155} + x^{154} + x^{153} -x^{152} + x^{149} + x^{148} + x^{147} -x^{146} + x^{144} -x^{141} -x^{140} -x^{137} +
	x^{136} + x^{135} + x^{134} -x^{133} -x^{132} -x^{131} -x^{130} -x^{129} -x^{128} + x^{127} + x^{126} + x^{125} +
	x^{124} + x^{123} + x^{122} + x^{121} -x^{119} -x^{118} -x^{117} -x^{116} + x^{115} + x^{114} + x^{111} + x^{110} -x^{109}
	-x^{108} -x^{107} -x^{104} + x^{103} + x^{102} -x^{101} -x^{100} -x^{98} -x^{97} -x^{95} -x^{94} -x^{93} +x^{90} + x^{89} -x^{88} + x^{87} + x^{85} -x^{84} + x^{82} -x^{81} -x^{80} -x^{79} -x^{77} + x^{76} + x^{73} -x^{71} +x^{70} + x^{69} + x^{68} -x^{67} -x^{65} + x^{64} -x^{62} + x^{61} -x^{59} -x^{58} + x^{57} + x^{56} + x^{55} -x^{53} -x^{52} -x^{49} + x^{47} + x^{46} -x^{45} + x^{44} -x^{41} + x^{39} + x^{37} + x^{36} -x^{35} -x^{33} -x^{32} + x^{31} +x^{30} -x^{29} -x^{24} -x^{22} -x^{20} -x^{18} + x^{17} + x^{14} + x^{13} + x^{12} + x^{10} + x^{9} -x^{8} -x^{7} -x^{6} -x^{4} + x^{3} + x^{2} -1)\cdot m\left ( x \right ) $,\\
	where $m(x)$ is the product of the  remained irreducible polynomials over $\mathbb{F}_3$. Thus $(x+1)^e-x^e-1=0$ has at least 773 solutions in $\mathbb{F}_{3^{773} }$ by Lemma \ref{l12}. And then from Lemma \ref{l14} $Q_3$, the ternary cyclic code  $\mathcal{C}_{(1, e)}$ is not optimal with respect to the Sphere Packing Bound.
	
	\vspace{0.8em}For convenience, in the following Lemmas \ref{l21}-\ref{l23} and Theorem \ref{p1}, we assume that $h$ is an integer with $3 \leq h \leq m-1$ and $m$  is an odd prime with
	
	(I) $m\geq 5$ and $2h\equiv3\left ( \bmod~m \right )$;  
	
	\noindent or
	
	(II) $m\equiv2$ $\left ( \bmod~3 \right )$ and $h=\frac{m+1}{3}$.
	\begin{lemma}\label{l21}
		For $m\geq 5$ prime, $e=3^h+13$ with $3\leq h \leq m-1$, we have $e \notin C_1$ and $|C_e|=m$.
	\end{lemma}
	
	{\bf Proof.} It's easy to see that $e \notin C_1$ since $e$ is even. Now from Lemma \ref{l13} we have $|C_e| \mid m$, thus $|C_e|=1$ or $|C_e|=m$ since $m$ is prime. 
	
	If $|C_e|=1$, then $2\left(3^h+13\right)\equiv0\left( \bmod~3^m-1 \right)$, i.e., $\left(3^m-1\right) \mid 2\left(3^h+13\right)$. Note that $2\left(3^h+13\right) \leq 3^m-1$, thus $3^m-1=2\left(3^h+13\right)$, i.e., $3^h\left(3^{m-h}-2\right)=27$ which implies $h=3$, and so $3^{m-h}-2=1$, which leads to $m-h=1$. So $m=4$ which is a contradiction with the assumption $m\geq 5$. Hence, $|C_e|=m.\hfill\Box$
	
	\begin{lemma}\label{l22}
		For $e=3^h+13$, the equation
		\begin{equation}\label{eqn-1}
			(x+1)^e+x^e+1=0
		\end{equation}
		has the unique solution $x=1$ in $\mathbb{F}_{3^m}$.
	\end{lemma}
	
	{\bf Proof.} It's easy to see that $x=1$ is the unique solution in $\mathbb{F}_3$ for (\ref{eqn-1}). Note that (\ref{eqn-1}) can be simplified as
	$$
		x^{3^h}(x^6+x^4-x+1)=x^6-x^5+x^2+1.
	$$
    Suppose that $\theta \in \mathbb{F}_{3^m} \setminus \mathbb{F}_{3} $ is a solution for (\ref{eqn-1}), then we have  
	$$
		\theta^{3^h}(\theta^6+\theta^4-\theta+1)=\theta^6-\theta^5+\theta^2+1.
	$$
	If $\theta^6+\theta^4-\theta+1=0$, then $\theta^6-\theta^5+\theta^2+1=0$. Thus we have  $\theta^5+\theta^4=\theta^2+\theta$, which means that $\theta^3(\theta+1) =\theta+1$. Hence $\theta^3=1$ or $\theta=-1$, which implies that $\theta \in  \mathbb{F}_{3}$. This contradicts with the assumption $\theta \in \mathbb{F }_{3^m}\setminus \mathbb{F}_{3}$. Hence $\theta^6+\theta^4-\theta+1 \ne 0$, and so
	\begin{equation}\label{eqn-2}
		\theta^{3^h}=\frac{\theta^6-\theta^5+\theta^2+1}{\theta^6+\theta^4-\theta+1} \triangleq \frac{f(\theta)}{g(\theta)}.
	\end{equation}
	
	(I) For $m\geq 5$ and $2h\equiv3\left ( \bmod~m \right )$, we have $2h-3=mk\left(k\in \mathbb{Z^{+}}\right)$. Note that $\theta ^{3^{m}}=\theta$, we obtain $\theta ^{3^{2h}}=\theta ^{3^{km+3}}=\theta^{27}$. Thus by taking $3^{h}$-th power on both sides of (\ref{eqn-2}), we have
		\begin{align}\label{eqn-3}
			\theta^{27}=\frac{f(\theta)^{3^h}}{g(\theta)^{3^h}}&=\frac{f(\theta)^6-f(\theta)^5 g(\theta)+f(\theta)^2 g(\theta)^4+g(\theta)^6}{f(\theta)^6+f(\theta)^4 g(\theta)^2-f(\theta)g(\theta)^5+g(\theta)^6} \notag \\
			& \triangleq \frac{F(\theta)}{G(\theta)}.
		\end{align}
	It then follows from (\ref{eqn-3}) that
	\begin{align}
		F(\theta)-\theta^{27} G(\theta) 
		=&\theta^{63} -\theta^{59} -\theta^{58} -\theta^{57} + \theta^{56} -\theta^{55}-\theta^{54}+\theta^{53} \notag \\
		&+ \theta^{52} -\theta^{51} + \theta^{50} + \theta^{49}-\theta^{48} -\theta^{45}+\theta^{44} -\theta^{41} \notag\\
		&+ \theta^{40} + \theta^{39} -\theta^{38} + \theta^{37}+ \theta^{35}-\theta^{34} -\theta^{32} + \theta^{31} \notag\\
		& + \theta^{29} -\theta^{28} -\theta^{26} + \theta^{25}-\theta^{24} -\theta^{23}+ \theta^{22} -\theta^{19} \notag\\
		& + \theta^{18} + \theta^{15} -\theta^{14}-\theta^{13}+ \theta^{12} -\theta^{11} -\theta^{10} + \theta^{9} \notag\\
		&+ \theta^{8} -\theta^{7} + \theta^{6} + \theta^{5} +\theta^{4} -1=0. \notag
		\end{align}
	Basing on Magma program, we know that the left-hand side of the above equation can be factorized into the product of the irreducible polynomials over $\mathbb{F}_3$ as follows,
\begin{align}\label{eqn-4}
	F(\theta)-\theta^{27} G(\theta) 
	=&(\theta -1)^{9}(\theta^3 + \theta^2 + \theta-1)^{4}(\theta^3 -\theta^2 -\theta -1)^{4}\notag \\
	&(\theta^{15} + \theta^{14}-\theta^{13} + \theta^{11} + \theta^{10} + \theta^8 + \theta^7 + \theta^6-\theta^5 + \theta -1)\notag \\
	&(\theta^{15} -\theta^{14} + \theta^{10} -\theta^9 -\theta^8 -\theta^7 -\theta^5 -\theta^4 + \theta^2 -\theta -1).
\end{align}
	Now from the prime $m \geq 5$ and Lemma \ref{l12}, we know that (\ref{eqn-4}) has no solutions in $\mathbb{F}_{3^m} \backslash \mathbb{F}_3$. Therefore, $x=1$ is the unique solution in $\mathbb{F}_{3^m}$ for (\ref{eqn-1}).
	
	(II) For  $m\equiv2\left(\bmod~3\right)$ and $h=\frac{m+1}{3}$. Note that $\theta^{3^{m}}=\theta$, we obtain $\theta^{3^{3h}}=\theta^{3^{m+1}}=\theta^3$. Thus by taking $3^{2h}$-th power on both sides of (\ref{eqn-2}), we have
		\begin{align}\label{eqn-5}
			\theta^{3}=\frac{F(\theta)^{3^h}}{G(\theta)^{3^h}}&=\frac{F(\theta)^6-F(\theta)^5 G(\theta)+F(\theta)^2 G(\theta)^4+G(\theta)^6}{F(\theta)^6+F(\theta)^4 G(\theta)^2-F(\theta)G(\theta)^5+G(\theta)^6} \notag \\
			& \triangleq \frac{S(\theta)}{T(\theta)}.
		\end{align}
	It then follows from (\ref{eqn-5}) that
	\begin{align*}
		S(\theta)-\theta^{3} T(\theta) 
	=&\theta^{219} -\theta^{216} + \theta^{207} + \theta^{205} -\theta^{204} + \theta^{203} -\theta^{202}+ \theta^{201}-\theta^{200} -\theta^{199}\\ 
	&-\theta^{198} -\theta^{196} -\theta^{194} +\theta^{190} -\theta^{187} + \theta^{186}+\theta^{185} -\theta^{183} -\theta^{182} -\theta^{181} \\
	&+ \theta^{180} + \theta^{178} -\theta^{175} + \theta^{174}+\theta^{173} + \theta^{171} +\theta^{169} -\theta^{168} + \theta^{166} -\theta^{165}\\ 
	&+ \theta^{163} -\theta^{162}+\theta^{161} + \theta^{160} + \theta^{158} + \theta^{157} + \theta^{155} + \theta^{154} + \theta^{152} + \theta^{151}\\
	&+\theta^{149} + \theta^{148} + \theta^{147} + \theta^{146} + \theta^{145} -\theta^{144} + \theta^{143} + \theta^{142}+\theta^{140} + \theta^{139} \\
	&+ \theta^{137} + \theta^{136} + \theta^{134} + \theta^{132} +\theta^{131} + \theta^{129} +\theta^{128} -\theta^{127} -\theta^{125} -\theta^{124}\\
	& + \theta^{121} -\theta^{120} + \theta^{119} + \theta^{118} +\theta^{117} -\theta^{115} -\theta^{111} -\theta^{110} + \theta^{109} + \theta^{108}\\
	& + \theta^{104} -\theta^{102}-\theta^{101} -\theta^{100} + \theta^{99} -\theta^{98} + \theta^{95} + \theta^{94} + \theta^{92} -\theta^{91} -\theta^{90}\\
	&-\theta^{88} -\theta^{87} -\theta^{85} -\theta^{83} -\theta^{82} -\theta^{80} -\theta^{79} -\theta^{77}-\theta^{76} + \theta^{75}-\theta^{74} \\
	&-\theta^{73} -\theta^{72} -\theta^{71} -\theta^{70} -\theta^{68} -\theta^{67}-\theta^{65} -\theta^{64} -\theta^{62}-\theta^{61} -\theta^{59}\\
	& -\theta^{58} + \theta^{57} -\theta^{56} + \theta^{54}-\theta^{53} + \theta^{51} -\theta^{50} -\theta^{48} -\theta^{46} -\theta^{45} + \theta^{44} \\
	&-\theta^{41} -\theta^{39}+\theta^{38} + \theta^{37} + \theta^{36} -\theta^{34} -\theta^{33} +\theta^{32} -\theta^{29} + \theta^{25} + \theta^{23}\\
	&+\theta^{21} + \theta^{20} + \theta^{19}-\theta^{18} + \theta^{17} -\theta^{16} + \theta^{15} -\theta^{14} -\theta^{12} + \theta^{3} -1=0.
\end{align*}
	Basing on Magma program, we know that the left-hand side of the above equation can be factorized into the product of the irreducible polynomials over $\mathbb{F}_3$ as follows,
	\begin{equation}\label{eqn-6}
		\begin{aligned}
			S(\theta)-\theta^{3}T(\theta)
		=&(\theta -1)^3(\theta^2 + 1)(\theta^2 + \theta -1)(\theta^2 -\theta-1)(\theta^4 + \theta^2 -\theta + 1)(\theta^4 + \theta^3 + \theta^2 + \theta\\
		& + 1)(\theta^4 -\theta^3 + \theta^2 + 1)(\theta^{10} -\theta^9 -\theta^4 + \theta^3+ \theta^2 + \theta -1)(\theta^{10} -\theta^9 -\theta^8 -\theta^7\\
		& + \theta^6 + \theta -1)(\theta^{10} -\theta^9 -\theta^8 -\theta^7 + \theta^6 + \theta^5 + \theta^4 -\theta^3 -\theta^2 -\theta + 1)\\
		&(\theta^{56} -\theta^{54}-\theta^{53} + \theta^{52}+ \theta^{51} + \theta^{49} + \theta^{48} + \theta^{47} + \theta^{46} + \theta^{44} + \theta^{43} + \theta^{42}+ \theta^{40} \\
		&-\theta^{39} -\theta^{38}-\theta^{36} + \theta^{34}-\theta^{32} -\theta^{29} + \theta^{28} -\theta^{27} -\theta^{24} + \theta^{22} -\theta^{20} -\theta^{18}\\
		& -\theta^{17}+ \theta^{16} + \theta^{14} + \theta^{13}+ \theta^{12} + \theta^{10} + \theta^9 + \theta^8 + \theta^7 + \theta^5 + \theta^4 -\theta^3 -\theta^2 + 1)\\
		&(\theta^{56} + \theta^{55} + \theta^{54} -\theta^{52} + \theta^{51} + \theta^{50} + \theta^{48} -\theta^{47} -\theta^{46} -\theta^{44} + \theta^{43} + \theta^{40} + \theta^{39} \\
		&-\theta^{38} -\theta^{36}+ \theta^{35} -\theta^{33}
		+ \theta^{30} -\theta^{29} + \theta^{28}-\theta^{27} + \theta^{26} -\theta^{25} -\theta^{22} -\theta^{21}\\
		&-\theta{^{19}}-\theta^{17} + \theta^{16} +\theta^{13} + \theta^{11}+ \theta^{10} -\theta^8 -\theta^7 -\theta^6 + \theta^5 + \theta^3 -\theta + 1)(\theta^{56}\\
		& -\theta^{55}+ \theta^{53} + \theta^{51} -\theta^{50} -\theta^{49} -\theta^{48} + \theta^{46} +\theta^{45} + \theta^{43} + \theta^{40} -\theta^{39} -\theta^{37}\\
		& -\theta^{35} -\theta^{34} -\theta^{31} + \theta^{30} -\theta^{29} + \theta^{28} -\theta^{27} + \theta^{26}-\theta^{23} + \theta^{21} -\theta^{20} -\theta^{18}\\
		& + \theta^{17} + \theta^{16} + \theta^{13} -\theta^{12} -\theta^{10} -\theta^9 + \theta^8 + \theta^6 + \theta^5 -\theta^4 + \theta^2 + \theta + 1).
	\end{aligned}
	\end{equation}
	Now from the prime $m \geq 5$ and Lemma \ref{l12}, we know that (\ref{eqn-6}) has no solutions in $\mathbb{F}_{3^m} \backslash \mathbb{F}_3$. Therefore, $x=1$ is the unique solution in $\mathbb{F}_{3^m}$ for (\ref{eqn-1}).$\hfill\Box$
	
	\begin{lemma}\label{l23}
		For $e=3^h+13$, the equation
		\begin{equation}\label{eqn-7}
			(x+1)^e-x^e-1=0
		\end{equation}
		has the unique solution $x=0$ in $\mathbb{F}_{3^m}$.
	\end{lemma}
	
	{\bf Proof.} It's easy to see that $x=0$ is the unique solution in $\mathbb{F}_3$ for (\ref{eqn-7}). Note that (\ref{eqn-7}) can be simplified as
	$$
		x^{3^h}(x^6-x^5+x^2+1)=-x^7-x^5+x^2-x.
	$$
	Suppose that $\theta \in \mathbb{F}_{3^m} \backslash \mathbb{F}_3$ is a solution for (\ref{eqn-7}), then we have
	$$
		\theta^{3^h}(\theta^6-\theta^5+\theta^2+1)=-\theta^7-\theta^5+\theta^2-\theta.
	$$
	If $\theta^6-\theta^5+\theta^2+1=0$, then $\theta^7-\theta^6+\theta^3+\theta=\theta \left(\theta^6-\theta^5+\theta^2+1 \right)=0$ and  $-\theta^7-\theta^5+\theta^2-\theta=0$. Thus we have $\theta^6+\theta^5=\theta^3+\theta^2$, which means that  $\theta^3(\theta+1)=\theta+1$. Hence $\theta^3=1$ or $\theta=-1$, which implies that $\theta \in  \mathbb{F}_{3}$. This contradicts with the assumption $\theta \in \mathbb{F }_{3^m}\setminus \mathbb{F}_{3}$. Hence $\theta^6-\theta^5+\theta^2+1\neq 0$, and so
	\begin{equation}\label{eqn-8}
		\theta^{3^h}=\frac{-\theta^7-\theta^5+\theta^2-\theta}{\theta^6-\theta^5+\theta^2+1} \triangleq \frac{f(\theta)}{g(\theta)}.
	\end{equation}
	
	(I) For $m\geq 5$ and $2h\equiv3\left ( \bmod~m \right )$. In the similar proof as that of (I) in Lemma \ref{l22}, by taking $3^h$-th power on both sides of (\ref{eqn-8}), we have
		\begin{align}\label{eqn-9}
			\theta^{27}=\frac{f(\theta)^{3^h}}{g(\theta)^{3^h}}&=\frac{-f(\theta)^7-f(\theta)^5 g(\theta)^2+f(\theta)^2 g(\theta)^5-f(\theta) g(\theta)^6}{f(\theta)^6 g(\theta)-f(\theta)^5 g(\theta)^2+f(\theta)^2 g(\theta)^5+g(\theta)^7}\notag \\
			& \triangleq \frac{F(\theta)}{G(\theta)}.
		\end{align}
	It then follows from (\ref{eqn-9}) that
	$$
	\begin{aligned}
		\theta^{27} G(\theta)-F(\theta) 
	=&\theta^{75} + \theta^{73} + \theta^{71} + \theta^{69} + \theta^{66} -\theta^{65} -\theta^{63}
	-\theta^{62} \\
	&-\theta^{61} + \theta^{60}+ \theta^{59} + \theta^{56} + \theta^{55} -\theta^{53}-\theta^{51} +\theta^{50} \\
	&-\theta^{49} + \theta^{48} -\theta^{47} -\theta^{46} 
	+ \theta^{43} +\theta^{42} + \theta^{41} -\theta^{40}\\ &-\theta^{39} + \theta^{37} + \theta^{36} -\theta^{35} 
	-\theta^{34} -\theta^{33} + \theta^{30} + \theta^{29}\\ &-\theta^{28} + \theta^{27} -\theta^{26} +\theta^{25} + \theta^{23} -\theta^{21} -\theta^{20} -\theta^{17} \\
	&-\theta^{16} + \theta^{15}+\theta^{14} + \theta^{13}+ \theta^{11} -\theta^{10} -\theta^{7} -\theta^{5} -\theta^{3} -\theta=0.
		\end{aligned}
	$$
	Basing on Magma program, we know that the left-hand side of the above equation can be factorized into the product of the irreducible polynomials over $\mathbb{F}_3$ as follows,
		\begin{align}\label{eqn-10}
			\theta^{27} G(\theta)-F(\theta)=&\theta(\theta + 1)(\theta -1)(\theta^3 -\theta -1)^4(\theta^3 + \theta^2 -1)^4\notag \\
		&(\theta^{12} + \theta^{10} + \theta^9 + \theta^7 -\theta^6 + \theta^5 + \theta^3 -\theta^2 -\theta + 1) \notag\\
		&(\theta^{12} -\theta^{10} + \theta^9 -\theta^8 + \theta^7 + \theta^4 -\theta^3 -\theta^2 + 1) \notag\\
		&(\theta^{12} -\theta^{10} -\theta^9+ \theta^8 + \theta^5 -\theta^4 + \theta^3 -\theta^2 + 1)\notag \\
		&(\theta^{12} -\theta^{11} -\theta^{10} + \theta^9 + \theta^7 -\theta^6 + \theta^5 + \theta^3 + \theta^2 + 1).
	\end{align}
	Now from the prime $m \geq 5$ and Lemma \ref{l12}, we know that (\ref{eqn-10}) has no solutions in $\mathbb{F}_{3^m} \backslash \mathbb{F}_3$. Therefore, $x=0$ is the unique solution in $\mathbb{F}_{3^m}$ for (\ref{eqn-7}).$\hfill\Box$
	
	(II) For  $m\equiv2$ $(\bmod~3)$ and $h=\frac{m+1}{3}$. In the similar proof as that of (II) in Lemma \ref{l22}, by taking $3^{2h}$-th power on both sides of (\ref{eqn-8}), we have
		\begin{align}\label{eqn-11}
		\theta^{3}=\frac{F(\theta)^{3^h}}{G(\theta)^{3^h}}&=\frac{-F(\theta)^7-F(\theta)^5 G(\theta)^2+F(\theta)^2 G(\theta)^5-F(\theta) G(\theta)^6}{F(\theta)^6 G(\theta)-F(\theta)^5 G(\theta)^2+F(\theta)^2 G(\theta)^5+G(\theta)^7}\notag \\
			& \triangleq \frac{S(\theta)}{T(\theta)}.
		\end{align}
	It then follows from (\ref{eqn-11}) that
	\begin{align*}
	\theta^{3} T(\theta)-S(\theta) =&\theta^{345} -\theta^{344} -\theta^{343} + \theta^{340} -\theta^{337} + \theta^{335} + \theta^{334} + \theta^{333} + \theta^{332} -\theta^{331}\\
	&-\theta^{329}-\theta^{328}-\theta^{326}-\theta^{323} -\theta^{322} -\theta^{320} -\theta^{318} + \theta^{316} -\theta^{315} + \theta^{314}\\
	&+\theta^{313} -\theta^{312} -\theta^{311} -\theta^{310} -\theta^{309} +\theta^{308} -\theta^{303} + \theta^{301} + \theta^{299} -\theta^{297}\\
	&+\theta^{296}-\theta^{295}-\theta^{294}+\theta^{293}+\theta^{292}-\theta^{291} + \theta^{290} -\theta^{289} +\theta^{286}+\theta^{285}\\
	&-\theta^{283} -\theta^{282} + \theta^{281} + \theta^{278} + \theta^{277} -\theta^{276} + \theta^{272} -\theta^{270} + \theta^{268}-\theta^{266}\\
	&-\theta^{264}+\theta^{262}+\theta^{260}-\theta^{259}-\theta^{258}+\theta^{257}-\theta^{256}-\theta^{255}+\theta^{254}-\theta^{253}\\
	&-\theta^{252} + \theta^{251} -\theta^{250} + \theta^{249}-\theta^{247} -\theta^{246} + \theta^{245} + \theta^{242} -\theta^{241}-\theta^{239}\\
	&-\theta^{236} -\theta^{235} -\theta^{233} -\theta^{232} + \theta^{230} -\theta^{229}-\theta^{228} + \theta^{226} + \theta^{225}-\theta^{224}\\
	&+\theta^{223}+\theta^{222}+\theta^{221} + \theta^{220} + \theta^{218}+\theta^{217} + \theta^{215} -\theta^{214} + \theta^{212}+\theta^{211}\\
	&+\theta^{210}-\theta^{209}-\theta^{208}-\theta^{207}+\theta^{206}-\theta^{205}-\theta^{204}+ \theta^{203} -\theta^{202}+\theta^{201}\\
	&+\theta^{199}-\theta^{198}-\theta^{197}-\theta^{195}+\theta^{194}+\theta^{193}+\theta^{191}+\theta^{190}+\theta^{189}-\theta^{188}\\
	&-\theta^{187}+\theta^{186}-\theta^{185}-\theta^{184}-\theta^{183}-\theta^{182}-\theta^{181}-\theta^{180}-\theta^{177}-\theta^{176}\\
	&-\theta^{175}+\theta^{173}-\theta^{171}-\theta^{170}-\theta^{169}-\theta^{166}-\theta^{165}-\theta^{164}-\theta^{163}-\theta^{162}\\
	&-\theta^{161}+\theta^{160}-\theta^{159}-\theta^{158}+\theta^{157}+\theta^{156}+\theta^{155}+\theta^{153}+\theta^{152}-\theta^{151}\\
	&-\theta^{149}-\theta^{148}+\theta^{147}+\theta^{145}-\theta^{144}+\theta^{143}-\theta^{142}-\theta^{141}+\theta^{140}-\theta^{139}\\
	&-\theta^{138}-\theta^{137}+\theta^{136}+\theta^{135}+\theta^{134}-\theta^{132}+\theta^{131}+\theta^{129}+\theta^{128}+\theta^{126}\\
	&+\theta^{125}+\theta^{124}+\theta^{123}-\theta^{122}+\theta^{121}+\theta^{120}-\theta^{118}-\theta^{117}+\theta^{116}-\theta^{114}\\
	&-\theta^{113}-\theta^{111}-\theta^{110}-\theta^{107}-\theta^{105}+\theta^{104}+\theta^{101}-\theta^{100}-\theta^{99}+\theta^{97}\\
	&-\theta^{96}+\theta^{95}-\theta^{94}-\theta^{93}+\theta^{92}-\theta^{91}-\theta^{90}+\theta^{89}-\theta^{88}-\theta^{87}+\theta^{86}\\
	&+\theta^{84}-\theta^{82}-\theta^{80}+\theta^{78}-\theta^{76}+\theta^{74}-\theta^{70}+\theta^{69}+\theta^{68}+\theta^{65}-\theta^{64}\\
	&-\theta^{63}+\theta^{61}+\theta^{60}-\theta^{57}+\theta^{56}-\theta^{55}+\theta^{54}+\theta^{53}-\theta^{52}-\theta^{51}+\theta^{50}\\
	&-\theta^{49}+\theta^{47}+\theta^{45}-\theta^{43}+\theta^{38}-\theta^{37}-\theta^{36}-\theta^{35}-\theta^{34}+\theta^{33}+\theta^{32}\\
	&-\theta^{31}+\theta^{30}-\theta^{28}-\theta^{26}-\theta^{24}-\theta^{23}-\theta^{20}-\theta^{18}-\theta^{17}-\theta^{15}+\theta^{14}\\
	&+\theta^{13}+\theta^{12}+\theta^{11}-\theta^{9}+\theta^{6}-\theta^{3}-\theta^{2}+\theta.
	\end{align*}
	Basing on Magma program, we know that the left-hand side of the above equation can be factorized into the product of the irreducible polynomials over $\mathbb{F}_3$ as follows,
	\begin{equation}\label{eqn-12}
		\begin{aligned}
			\theta^{3}T(\theta)-S(\theta)
			=&\theta(\theta^4 + \theta^2 + \theta + 1)(\theta^4 + \theta^3 -\theta + 1)(\theta^4 + \theta^3 + \theta^2 + 1)(\theta^4 -\theta^3 + \theta + 1)(\theta^4 \\
			&-\theta^3 + \theta^2 -\theta + 1)(\theta^{20} + \theta^{19} -\theta^{17} + \theta^{16} -\theta^{14} + \theta^{13} -\theta^{10} -\theta^8 +\theta^7+ \theta^6 \\
			&-\theta^2 -\theta + 1)(\theta^{20} -\theta^{19} -\theta^{18} + \theta^{14} + \theta^{13} -\theta^{12} -\theta^{10} + \theta^7-\theta^6 + \theta^4-\theta^3\\
			& + \theta + 1)(\theta^{28} + \theta^{27} + \theta^{26} + \theta^{25} + \theta^{23} -\theta^{22} + \theta^{21} -\theta^{20} -\theta^{19}-\theta^{18} -\theta^{17}\\
			& + \theta^{16} + \theta^{12} -\theta^{11} -\theta^{10} -\theta^9 -\theta^8 + \theta^7 -\theta^6 + \theta^5 + \theta^3 + \theta^2 + \theta + 1)(\theta^{28}\\
			&+\theta^{27}-\theta^{26}-\theta^{25}-\theta^{24}+\theta^{21}+\theta^{19}+\theta^{18}-\theta^{16}-\theta^{15}-\theta^{14}-\theta^{13}-\theta^{12}+\theta^{10}\\
			&+\theta^9 + \theta^7 -\theta^4-\theta^3 -\theta^2 + \theta + 1)(\theta^{28} -\theta^{27} -\theta^{26} -\theta^{25}-\theta^{24} + \theta^{23} + \theta^{22}\\
			&+\theta^{20}+\theta^{16}-\theta^{15}-\theta^{14}-\theta^{13}+\theta^{12}+\theta^8 + \theta^6+\theta^5 -\theta^4 -\theta^3-\theta^2-\theta+1)\\
			&(\theta^{200}+\theta^{199}+\theta^{198}+\theta^{196}+\theta^{195}-\theta^{193}-\theta^{191}-\theta^{190}-\theta^{189}-\theta^{188}-\theta^{186}\\
			&-\theta^{185}+\theta^{184}-\theta^{183}+\theta^{182}+\theta^{181}+\theta^{180}-\theta^{179}+\theta^{178}-\theta^{176}-\theta^{175}-\theta^{173}\\
			&+\theta^{171}+\theta^{169}+\theta^{168} +\theta^{166}-\theta^{165}-\theta^{163}-\theta^{162}+\theta^{159}-\theta^{158}-\theta^{157}-\theta^{156}\\
			&-\theta^{150}-\theta^{149}+\theta^{148}+\theta^{147}-\theta^{145}+\theta^{143}+\theta^{142}+\theta^{141}+\theta^{139}+\theta^{136}-\theta^{135}\\
			& -\theta^{134} -\theta^{133} -\theta^{132}-\theta^{131}-\theta^{130} -\theta^{129} + \theta^{127} -\theta^{126} + \theta^{125}+\theta^{123}+\theta^{122}\\
			&+\theta^{121}+\theta^{118}+\theta^{117}+\theta^{115}-\theta^{114} -\theta^{111}-\theta^{109}-\theta^{108}+\theta^{107}+\theta^{106}-\theta^{105}\\
			&+\theta^{101}+\theta^{100}+\theta^{99}-\theta^{95}+\theta^{94}+\theta^{93}-\theta^{92}-\theta^{91}-\theta^{89}-\theta^{86}+\theta^{85}+\theta^{83}\\
			&+\theta^{82}+\theta^{79}+\theta^{78}+\theta^{77}+\theta^{75}-\theta^{74}+\theta^{73}-\theta^{71}-\theta^{70}-\theta^{69}-\theta^{68}-\theta^{67}-\theta^{66}\\
			&-\theta^{65}+\theta^{64}+\theta^{61}+\theta^{59}+\theta^{58}+\theta^{57}-\theta^{55}+\theta^{53}+\theta^{52}-\theta^{51}-\theta^{50}-\theta^{44}-\theta^{43}\\
			&-\theta^{42}+\theta^{41}-\theta^{38}-\theta^{37}-\theta^{35}+\theta^{34}+\theta^{32}+\theta^{31}+\theta^{29}-\theta^{27} -\theta^{25}-\theta^{24}+\theta^{22}\\
			&-\theta^{21}+\theta^{20}+\theta^{19}+\theta^{18}-\theta^{17}+\theta^{16}-\theta^{15}-\theta^{14}-\theta^{12}-\theta^{11}-\theta^{10}-\theta^9-\theta^7\\
			&+\theta^5+ \theta^4 + \theta^2 + \theta + 1).
		\end{aligned}
	\end{equation}
	Now from the prime $m \geq 5$ and Lemma \ref{l12}, we know that (\ref{eqn-12}) has no solutions in $\mathbb{F}_{3^m} \backslash \mathbb{F}_3$. Therefore, $x=0$ is the unique solution in $\mathbb{F}_{3^m}$ for (\ref{eqn-7}).$\hfill\Box$
	
	According to the above Lemmas \ref{l21}-\ref{l23}, we obtain an incomplete answer for the 8th problem as follows, which is also a sufficient condition for that $\mathcal{C}_{(1, e)}$ in the 8th problem is optimal with respect to the Sphere Packing Bound. 
	\begin{theorem}\label{p1}
		For $e=3^h+13$, the ternary cyclic code $\mathcal{C}_{(1, e)}$ is an optimal ternary cyclic code with parameters$$[3^m-1,3^m-1-2 m, 4].$$
	\end{theorem}

\section{The second class of optimal ternary cyclic codes with minimum distance four}\label{h3}

\subsection{The second class of optimal ternary cyclic codes $\mathcal{C}_{(1, e)}$ }
	In this section, by empolying the properties of finite fields, we obtain a class of optimal ternary cyclic codes $\mathcal{C}_{(1, e)}$ with parameters $[3^m-1,3^m-1-2 m, 4]$.
	
	The following Lemma \ref{g1} and Corollary \ref{j1} are necessary.
	
		\begin{lemma}\label{g1}
		For any positive integers $s$, $n$ and $p$ prime  with $\operatorname{gcd}\left(p^s-1, n\right)=1$. If $t\in \mathbb{F}_{p^s}^{*}$, then there exists some $\beta \in \mathbb{F}_{p^s}^{*}$ such that  $t=\beta^n$.
		\end{lemma}
		
			{\bf Proof.} Assume that $\alpha$ is a generator of $\mathbb{F}_{p^s}^{*}$ and $t=\alpha ^{k}\left ( 0 \leq k \leq p^s-2  \right )$. Note that the equation $nx\equiv k\left ( \bmod~{p^s-1} \right)$ has an unique solution $r$ from $\operatorname{gcd}\left(p^s-1, n\right)=1$. Now set $\beta=\alpha^{r}$, it's easy to see that 
		\begin{equation*}
			t=\alpha ^{k}=\alpha ^{nr}=\left ( \alpha ^{r}  \right )^{n }=\beta^{n}.
		\end{equation*}
		This completes the proof.$\hfill\Box$		
		
		\begin{corollary}\label{j1}
			For any positive integers $m$ and $n$.
			
		(1) If $t\in \mathbb{F}_{3^{m} }^{*}$, then there exists some $\beta \in \mathbb{F}_{3^{m} }^{*}$ such that  $t=\beta^{3^n}$;
		
		(2) If $t \in \mathbb{F}_{3^{m}}^{*}\setminus \left \{ -1 \right \} $, then there exists some $\beta$ and $\theta \in \mathbb{F}_{3^m}^{*}$  such that $t=\beta^{3^n}$, $t+1=\theta^{3^n}$ and $\theta=\beta+1$.
			
		\end{corollary}
		
	\begin{theorem}\label{g3}
		For any positive integers $m$, $e$ and $n$ with $1\leq e\leq 3^{m}-2$ and $3^{n}e \equiv 2\left(\bmod~3^m-1\right)$, $\mathcal{C}_{(1, e)}$ is an optimal ternary cyclic code with parameters $$\left[3^m-1,3^m-1-2 m, 4\right].$$
	\end{theorem}
	
	{\bf Proof.} It's easy to see that $e$ is even, $e \notin C_1$ and $\operatorname{gcd}\left(e, 3^m-1\right)\mid 2$ since $3^{n}e \equiv 2\left(\bmod~3^m-1\right)$. 
	
	(1) Note that $2\mid\operatorname{gcd}\left(e, 3^m-1\right)$, we have $\operatorname{gcd}\left(e, 3^m-1\right)=2$, which means that $\left|C_e\right|=m$ by Lemma \ref{l11}. 
	
	(2) It's easy to see that $x=1$ is the unique solution in $\mathbb{F}_3$ for $(x+1)^e+x^e+1=0$. If $x_0 \in \mathbb{F }_{3^m}\setminus \mathbb{F}_{3} $ is a solution for $(x+1)^e+x^e+1=0$, then $x_0, x_0+1\in \mathbb{F}_{3^{m} }^{*}$ and 
	\begin{equation}\label{eqn-100}
	(x_0+1)^e+x_0^e+1=0.
	\end{equation}
	 There exists some $\theta$ and $\beta \in \mathbb{F}_{3^m}^{*}$  such that $x_0+1=\theta^{3^n}$, $x_0=\beta^{3^n}$ and $\theta=\beta+1$ by Corollary \ref{j1}, which implies that (\ref{eqn-100}) is equivalent to
	$$
	\theta ^{3^{n}e } +\beta ^{3^{n}e}+1=0.
	$$ 
	According to $3^n e \equiv 2~(\bmod~3^m-1)$ and $\theta ^{3^{m}-1 } =\beta ^{3^{m}-1}=1$, the above equation can be reduced to
	\begin{equation}\label{eqn-52}
		\theta^2+\beta^2+1=0.
	\end{equation}
	Now from (\ref{eqn-52}) and $\theta=\beta+1$, we have $\beta=1$, and so $x_0=\beta^{3^n}=1 \in \mathbb{F}_3$, which is a contradiction. Thus, $(x+1)^e+x^e+1=0$ has the unique solution $x=1$ in $\mathbb{F}_{3^m}$. 
	
	(3) In the similar proof as that of (2) in Theorem \ref{g3}, the equation
	$$
	(x+1)^e-x^e-1=0
	$$
	can be reduced to
	\begin{equation}\label{eqn-53}
		\theta^2-\beta^2-1=0.
	\end{equation}
	Now from (\ref{eqn-53}) and $\theta=\beta+1$, we have $\beta=0$. Thus, $(x+1)^e-x^e-1=0$ has the unique solution $x=0$ in $\mathbb{F}_{3^m}$.
	
	From the above and Lemma \ref{l14}, $\mathcal{C}_{(1, e)}$ is an optimal ternary cyclic code with parameters $\left[3^m-1,3^m-1-2 m, 4\right].$$\hfill\Box$
	
	\begin{lemma}\label{g2}
		For any positive integers $m$, $e$ and $n$ with $m>2$, $3^n e \equiv 4~(\bmod~3^m-1)$ and $1\leq e\leq 3^{m}-2$, we have $e \notin C_1$ and $\left|C_e\right|=m$.
	\end{lemma}
	
	{\bf Proof.} It's easy to see that $e$ is even, $e \notin C_1$ and $\operatorname{gcd}\left(e, 3^m-1\right)\mid4$ since $3^n e \equiv 4\left(\bmod~3^m-1\right)$. We have the following two cases depending on $m$ odd or even, respectively. 
	
	{\bf Case I.}
	For $m$ odd, we have $4\nmid \left ( 3^{m}-1  \right )$, then  $\operatorname{gcd}\left(e, 3^m-1\right)=2$, which means that $\left|C_e\right|=m$ from Lemma \ref{l11}. 
	
	{\bf Case II.} 
	For $m$ even, if there are two integers $i$, $j$ with $0 \leq i<j \leq m-1$ and $3^i \cdot e \equiv 3^j \cdot e\left(\bmod~ 3^m-1\right)$, then we have $1 \leq j-i \leq m-1$ and 
	$
	\left(3^m-1\right) \mid\left(3^{j-i}-1\right) \cdot e,
	$
	which means that $$\operatorname{gcd}\left(\left(3^{j-i}-1\right)\cdot e ,3^m-1\right)=3^{m}-1.$$
	Note that $\operatorname{gcd}(j-i, m)\leq \frac{m}{2}$, it can be easily checked that
	$$
	\operatorname{gcd}\left(3^{j-i}-1,3^m-1\right)=3^{\operatorname{gcd}(j-i, m)}-1 \leq 3^{\frac{m}{2}}-1.
	$$
	Especially, for $m>2$, we have
    \begin{align*}
	3^m-1=\operatorname{gcd}\left(3^m-1,\left(3^{j-i}-1\right) \cdot e\right) &\leq \operatorname{gcd}\left (3^m-1,e  \right ) \cdot \operatorname{gcd}\left (3^m-1,3^{j-i}-1  \right )\notag\\
	&\leq 4\left(3^{\frac{m}{2}}-1\right) \notag \\
	&<3^m-1,
	\end{align*}
	which is a contradiction. Therefore, $3^i \cdot e \not \equiv 3^j \cdot e \left(\bmod~3^m-1\right)$ for any $0 \leq i<$ $j \leq m-1$, which means that $\left|C_e\right|=m$.$\hfill\Box$
	
	\begin{theorem}\label{g4}
		For any positive integers $m$, $e$ and $n$ with $m>2$ odd, $3^n e \equiv 4~(\bmod~3^m-1)$ and $1\leq e\leq 3^{m}-2$, $\mathcal{C}_{(1, e)}$ is an optimal ternary cyclic code with parameters$$\left[3^m-1,3^m-1-2 m, 4\right].$$
	\end{theorem}
	
	{\bf Proof.} (1) We know that $e$ is even since $3^n e \equiv 4~(\bmod~3^m-1)$, and then $e \notin C_1$, $\left|C_e\right|=m$ by Lemma \ref{g2}. 
	
	(2) It's easy to see that $x=1$ is the unique solution in $\mathbb{F}_3$ for $(x+1)^e+x^e+1=0$. If $x_0 \in \mathbb{F }_{3^m}\setminus \mathbb{F}_{3}$ is a solution for $(x+1)^e+x^e+1=0$, then $x_0, x_0+1\in \mathbb{F}_{3^{m} }^{*}$ and  
		\begin{equation}\label{eqn-101}
		(x_0+1)^e+x_0^e+1=0.
	\end{equation}
	 There exists some $\theta$ and $\beta \in \mathbb{F}_{3^m}^{*}$  such that $x_0+1=\theta^{3^n}$, $x_0=\beta^{3^n}$ and $\theta=\beta+1$ by Corollary \ref{j1}, which implies that (\ref{eqn-101}) is equivalent to
	$$
	\theta ^{3^{n}e } +\beta ^{3^{n}e}+1=0.
	$$ 
	Now from $3^n e \equiv 4~(\bmod~3^m-1)$ and $\theta ^{3^{m}-1 } =\beta ^{3^{m}-1}=1$, the above equation can be reduced to
	\begin{equation}\label{eqn-56}
		\theta^4+\beta^4+1=0.
	\end{equation}
	From $\theta=\beta+1$ and (\ref{eqn-56}), we have
	\begin{equation}\label{eqn-70}
		\beta^4-\beta^3-\beta+1=(\beta-1)^4=0.
	\end{equation}
	It's easy to show that (\ref{eqn-70}) has the unique solution $\beta=1$  and so $x_0=\beta^{3^n}=1 \in \mathbb{F}_3$, which is a contradiction. Thus $(x+1)^e+x^e+1=0$ has the unique solution $x=1$ in $\mathbb{F}_{3^m}$. 
	
	(3) In the similar proof as that of (2) in Theorem \ref{g4}, the equation
	$$
	(x+1)^e-x^e-1=0
	$$
	can be simplified to
	\begin{equation}\label{eqn-57}
		\theta^4-\beta^4-1=0.
	\end{equation}
	From $\theta=\beta+1$ and (\ref{eqn-57}), we have 
	\begin{equation}\label{eqn-71}
		\beta^3+\beta=\beta \left ( \beta ^{2}+1\right )=0.
	\end{equation}
	For $m$ odd, we obtain that (\ref{eqn-71}) has the unique solution $\beta=0$ in $\mathbb{F}_{3^m}$. And so $(x+1)^e-x^e-1=0$ has the unique solution $x=0$ in $\mathbb{F}_{3^m}$.
	
	From the above and Lemma \ref{l14}, $\mathcal{C}_{(1, e)}$ is an optimal ternary cyclic code with parameters $\left[3^m-1,3^m-1-2 m, 4\right].$$\hfill\Box$

\subsection{The equivalence of the optimal ternary cyclic codes $\mathcal{C}_{(1, e)}$ }

	The following Table 1 is about the known results for Ding and Helleseth's 8th open problem, and Table 2 is about the results in our Theorem 3.1. 
	\begin{table}[htbp]
		\centering
		\begin{threeparttable}
			\scriptsize  
			\renewcommand\arraystretch{1.1}
			\setlength\tabcolsep{0.5pt}
			\setlength{\abovecaptionskip}{3pt}
			\setlength{\belowcaptionskip}{0pt}
			\caption{{\footnotesize Known results of the Ding and Helleseth's 8th open problem}}
			
			{\begin{tabular}{|p{6em}|p{8em}|p{20em}|p{6em}|}  
					\hline
					\centering
					Type & \quad ~~~~~~~e & \quad~~~~~~~~~~~~~~~~~~~~Conditions &\quad~~~ Ref.\\
					
					\hline
					\quad ~~~~~$1$ 
					&\quad ~~~~$3^h+13$
					&\quad ~~~~~~~$h=\frac{m-1}{2}$, $m\geq 7$ is prime   
					&\quad ~~~~[12] \\
					
					\hline
					\quad ~~~~~$2$ 
					&\quad ~~~~$3^h+13$
					&\quad ~~~~~~~$h=\frac{m+1}{2}$, $m\geq 5$ is prime 
					&\quad ~~~~[14]\\
					
					\hline
					
			\end{tabular}}
		\end{threeparttable}
	\end{table}
	
	\begin{table}[htbp]
		\centering
		\begin{threeparttable}
			\scriptsize  
			\renewcommand\arraystretch{1.1}
			\setlength\tabcolsep{0.5pt}
			\setlength{\abovecaptionskip}{3pt}
			\setlength{\belowcaptionskip}{0pt}
			\caption{{\footnotesize The results in Theorem 3.1 }}
			
			
			{\begin{tabular}{|p{6em}|p{8em}|p{20em}|p{10em}|}  
					\hline
					\centering
					Type & \quad ~~~~~~~e & \quad~~~~~~~~~~~~~~~~~~~~Conditions &\quad~~~~~ Theorem\\
					
					\hline
					\quad ~~~~~$3$ 
					&\quad ~~~~$3^h+13$
					&\quad ~~~~~~~~$h=\frac{m+3}{2}$, $m\geq 5$ is prime 
					&\quad~~Theorem 3.1(I)\\
					
					\cline{1-4}
					\quad ~~~~~$4$ 
					&\quad ~~~~$3^h+13$
					&\quad ~~ $h=\frac{m+1}{3}$, $m\equiv2$ $\left ( \bmod~3 \right )$ is prime   
					&\quad~~Theorem 3.1(II) \\
					
					\hline						
			\end{tabular}}
			
		\end{threeparttable}
	\end{table}

	Let $u$ and $v$(or $v'$) be not in the same $p$-cyclotomic cost, in order to check that $\mathcal{C}_{(u, v)}$ is not equivalent to $\mathcal{C}_{(u, v')}$, the following Lemma 4.3 is needed.
\begin{lemma}\cite[Theorem 3.7]{A21}\label{l15}
		Let $v$ and $v'$ be in the same p-cyclotomic cost, then the cyclic code $\mathcal{C}_{(u, v)}$ is equivalent to $\mathcal{C}_{(u, v')}$.
\end{lemma}

	According to Lemma \ref{l15}, we can get Corollaries 4.2-4.4 immediately. Corollaries 4.2-4.3 imply that $\mathcal{C}_{(1, e)}$ in Table 2 is not equivalent to any known codes in Table 1. Corollary 4.4 implies that $\mathcal{C}_{(1, e)}$ in Theorem 4.1 or Theorem 4.2 is not equivalent to any known codes.
	
\vspace{1em} \noindent {\bf Corollary 4.2 }
		Let the prime $m > 7$ and the integer $e=3^{\frac{m+3}{2}}+13$, then the optimal ternary cyclic code $\mathcal{C}_{(1, e)}$ is not equivalent to $\mathcal{C}_{(1, e')}$, where $e'=3^{\frac{m-1}{2}}+13$ or $e'=3^{\frac{m+1}{2}}+13$.

 		
 		
 
\vspace{0.6em}

	{\bf Proof.}  First, we prove that $\mathcal{C}_{(1, e)}$ is not equivalent to $\mathcal{C}_{(1, e')}$, where $e'=3^{\frac{m-1}{2}}+13$, which only need to prove that $3^{\frac{m+3}{2}}+13$ and $3^{\frac{m-1}{2}}+13$ are not in the same $p$-cyclotomic cost. If there are two integers $i$, $j$ satisfying that $0 \leq i<j \leq m-1$ and
	\begin{equation}\label{eqn-1017}
	3^i \cdot (3^{\frac{m+3}{2}}+13) \equiv 3^j \cdot (3^{\frac{m-1}{2}}+13) ~(\bmod~3^m-1),
	\end{equation}
	then  (\ref{eqn-1017}) is equivalent to
	$$
	\begin{aligned}
		3^k \cdot(3^{\frac{m+3}{2}}+13) \equiv 3^{\frac{m-1}{2}}+13~(\bmod~3^m-1),
	\end{aligned}
	$$
	where $1 \leq k=j-i \leq m-1$. Note that $\frac{m-3}{2}-2 >0$ and
		$$
		\begin{aligned}
			3^{\frac{m+3}{2}}+13=3^{\frac{m+3}{2}}-1+14&=(3-1)\cdot(3^{\frac{m+1}{2}}+3^{\frac{m-1}{2}}+\cdots+3^2+3^1+3^0)+3^2+3^1+2\cdot3^0\\
		&=3^{\frac{m+3}{2}}+0\cdot3^{\frac{m+1}{2}}+\cdots+0\cdot3^3+3^2+3^1+ 3^0.
		\end{aligned}
		$$
	Thus, $3^{\frac{m+3}{2}}+13$ can be regarded as a 3-adic vector with length $m$ as follows,
	$$
	\begin{aligned}
		\alpha=(c_{m-1}, \cdots, c_0)=(0, \cdots, 0,1,\underbrace{0, \cdots,0}_\frac{m-3}{2},1,1,1).
	\end{aligned}
	$$
	Then $3^k \cdot(3^{\frac{m+3}{2}}+13) ~(\bmod~3^m-1)$ can be regarded as a 3-adic vector with length $m$ as follows,
	$$
	\beta=(c_{m-1-k}, \cdots, c_0, c_{m-1}, \cdots, c_{m-k}).
	$$
	Similarly, $3^{\frac{m-1}{2}}+13$ can be regarded as a 3-adic vector with length $m$ as follows,
	$$
	\begin{aligned}
		\gamma=(0, \cdots, 0,1,\underbrace{0, \cdots,0}_{\frac{m-3}{2}-2},1,1,1).
	\end{aligned}
	$$
	Now from $1 \leq k\leq m-1$, we know that $\beta \neq \gamma$, which means that
	$$
	\begin{aligned}
		3^k \cdot(3^{\frac{m+3}{2}}+13) \not \equiv  3^{\frac{m-1}{2}}+13 ~(\bmod~3^m-1),
	\end{aligned}
	$$
	namely, $3^{\frac{m+3}{2}}+13$ and $3^{\frac{m-1}{2}}+13$ are not in the same $p$-cyclotomic cost. From Lemma 4.3, we know that $\mathcal{C}_{(1, e)}$ is not equivalent to $\mathcal{C}_{(1, e')}$, where $e'=3^{\frac{m-1}{2}}+13$. 


For the case $e'=3^{\frac{m+1}{2}}+13$, in the similar proof, we also know that
$\mathcal{C}_{(1, e)}$ is not equivalent to $\mathcal{C}_{(1, e')}$. 

$\hfill\Box$

\vspace{0.8em} \noindent {\bf Corollary 4.3} Let the prime $m\equiv2$ $\left ( \bmod~3 \right )$ and the integer $e=3^{\frac{m+1}{3}}+13$, then the optimal ternary cyclic code $\mathcal{C}_{(1, e)}$ is not equivalent to $\mathcal{C}_{(1, e')}$, where $e'=3^{\frac{m-1}{2}}+13$ or $e'=3^{\frac{m+1}{2}}+13$.

	

\vspace{0.8em} \noindent {\bf Corollary 4.4} For any positive integers $m$, $e$ and $n$ with $1\leq e\leq 3^{m}-2$.

\vspace{0.4em} \noindent {~~(1)} If $3^{n}e \equiv 2\left(\bmod~3^m-1\right)$, then $\mathcal{C}_{(1, e)}$ is equivalent to $\mathcal{C}_{(1, e')}$, where $e'=2\cdot 3^{m-n}$.

\vspace{0.4em} \noindent {~~(2)}  If $3^{n}e \equiv 4 \left(\bmod~3^m-1\right)$ with $m>2$ odd, then $\mathcal{C}_{(1, e)}$ is equivalent to $\mathcal{C}_{(1, e')}$, where $e'=4\cdot 3^{m-n}$.



	\section{Conclusions}\label{h4}
	In this manuscript, we first give a counterexample for the 8th problem proposed by Ding and Helleseth \cite{A11}. Secondly, basing on properties and polynomials over finite fields, we obtain two sufficient conditions for the ternary cyclic code $\mathcal{C}_{(1, e)}$ optimal with respect to the Sphere Packing Bound as follows. 

	(1) $e=3^h+13$, $m\geq 5$ is prime and $2h\equiv3\left ( \bmod~m \right )$, or $m\equiv2$ $(\bmod~3)$ is prime and $h=\frac{m+1}{3}$;
	
	(2) $3^n e \equiv 2~(\bmod~3^m-1)$ with $m>2$, or $3^n e \equiv 4~(\bmod~3^m-1)$ with $m>2$ odd, where $e, n$ are positive integers and $1\leq e\leq 3^{m}-2$.
	
	Finally, it's easy to see that (1) is just an incomplete answer for the 8th problem proposed by Ding and Helleseth \cite{A11}, and show that these codes are not equivalent to any known codes.

\end{document}